\documentclass[10pt]{IEEEtran}
\usepackage{graphicx}
\usepackage{epsfig}
\usepackage{amsmath,amsfonts,dsfont,amssymb,stmaryrd}

\newtheorem{definition}{Definition}
\newtheorem{theorem}{Theorem}
\newtheorem{prop}{Proposition}
\newtheorem{lemma}{Lemma}
\newtheorem{corol}{Corollary}

\newcommand{\ww}{\widetilde{W}}

\newcommand{\mZ}{\mathbb{Z}}

\newcommand{\Sc}{\mathcal{S}}

\newcommand{\X}{\mathcal{X}}
\newcommand{\Y}{\mathcal{Y}}

\newcommand{\A}{\mathcal{A}}

\begin{document}

\title{Universal A Posteriori Metrics Game}

\author{
\begin{tabular}{c c}
    \begin{minipage}{0.5\linewidth}
        \begin{center}
        Emmanuel Abbe \\
                 LCM, EPFL \\
                 Lausanne, 1015, Switzerland \\
                emmanuel.abbe@epfl.ch \\
        \end{center}
    \end{minipage}
    &
    \begin{minipage}{0.5\linewidth}
        \vspace{-2pt}
        \begin{center}
        Rethnakaran Pulikkoonattu \thanks{R.Pulikkoonattu was with EPFL. He has been with Broadcom since October 2009.} \\
              Broadcom Inc,\\
             Irvine, CA, USA, 92604 \\
              rethna@broadcom.com \\
        \end{center}
    \end{minipage}
    \\
\end{tabular}
}

\maketitle



\begin{abstract}
Over binary input channels, uniform distribution is a universal prior, in the sense that it allows to
maximize the worst case mutual information over all binary input channels, ensuring at least 94.2\% of the capacity.
In this paper, we address a similar question, but with respect to a universal generalized linear decoder.
We look for the best collection of finitely many a posteriori metrics, to maximize the worst case mismatched
mutual information achieved by decoding with these metrics (instead of an optimal decoder such as the Maximum Likelihood (ML) tuned to the
true channel). It is shown that for binary input and output channels, two metrics suffice to actually achieve the same
performance as an optimal decoder.
In particular, this implies that
there exist a decoder
which is generalized linear
and achieves at
least $94.2\%$ of the compound capacity on any compound set, without the
knowledge of the underlying set.
\end{abstract}

\section{Introduction}

We consider the problem where a communication system is to be designed without
explicit knowledge of the channel. Here, neither the transmitter nor the receiver
have access to the exact channel law. The goal is to devise a single
coding strategy which will ensure reliable communication over the
unknown channel picked for transmission. 
We assume i.i.d. realizations of the
channel at each channel use, i.e., we are interested in a universal coding framework for communicating over discrete memoryless channels (DMC).
In this paper, we present results for DMC's with binary input and output alphabets, which we refer to as binary memoryless channels (BMC).
Our goal is to design decoders which have a linear structure and which entails
reliable communication at the largest possible rate in this setting.
In the next section, we revise in more details the notion of universality
and linearity for DMC's, and their attribute. We will then formulate our
problem as a game where the decoder has to pick the decoding metrics, i.e., a
generalized linear decoder, before nature select a channel for communication.
\subsection{Universality}
If the channel over which communication takes place is unknown at
both the transmitter and the receiver but belongs to a set of DMC's $\mathcal{S}$, then we
are in the setting of \textit{compound channels}.
Let us denote by $\X$ the input alphabet and $\Y$ the output alphabet of any DMC in $\mathcal{S}$.
The objective is to design a code (i.e., an encoder and decoder pair) which will provide a mean for reliable
communication, independently of which $W\in \mathcal{S}$ is picked up (by nature) for transmission.
The compound channel problem has been extensively studied in the literature such as \cite{paper:blackwell},\cite{paper:cover:broadcast},\cite{paper:Csiszer-Narayan1995},\cite{paper:compound:Dobrushin},\cite{paper:feder-lapodoth},\cite{paper:Lapidoth-Narayan1998} and \cite{paper:lapidoth-telatar1998}.

The highest achievable rate, known as the \textit{compound capacity} $C(\Sc)$ of a set $\mathcal{S}$ of
channels is established in \cite{paper:blackwell} and is given by:
\begin{eqnarray}\label{bbt}
C(\mathcal{S}) \triangleq \underset{P}{\max}\underset{W\in \mathcal{S}}{\inf} I(P,W)
\end{eqnarray}
where the maximization is over all possible probability distributions $P$ on $\mathcal{X}$,
and the infimum is performed over all the channels $W$ in the compound set $\mathcal{S}$.


In \cite{paper:blackwell}, a decoder that maximizes a uniform mixture of likelihoods, over a dense set of
possible channels is proposed as a universal decoder.
In the literature, a decoder which allows us to achieve the same random coding
exponent (without the knowledge of true channel) as the maximum likelihood (ML)
decoder tuned to the true channel is called a {\it universal decoder}.
The maximum mutual information (MMI) decoder introduced in \cite{paper:Goppa} is a universal decoder. The MMI
decoder computes the empirical mutual information (EMI) between a given received output
and each codewords in the codebook, and declares the element with the highest EMI score as the
sent codeword. There has been a number of other universal decoders proposed in the literature,
such as Lempell-Ziv (LZ) based algorithm \cite{paper:ziv}, and the merged
likelihood decoder \cite{paper:feder-lapodoth}.
The MMI decoder has another interesting feature: it does not even require the
knowledge of the compound set to be defined.
In that sense, the MMI decoder is a ``doubly universal'' decoder. However, practical use of MMI decoders are
voided by complexity considerations, and similarly for other existing universal decoders.
Note that in this paper, we are primarily concerned with the achievable rate rather than error exponent.


\subsection{Linearity}
A linear (or additive) decoder is defined to have the following structure. Upon receiving
each $n-$symbol output $y$, the decoder computes a score (decoding metric) $d^{n}\left(x_m,y\right)$
for each codeword $x_m, m=1,2,\ldots 2^{nR}$ and declares the codeword with the
highest score as estimate of the sent codeword(ties are resolved arbitrarily).
Moreover, the $n-$symbol decoding metric has
the following additive structure.
\begin{equation*}
d^{n}\left(x,y\right)=\sum_{i=1}^{n}{d\left(x(i),y(i)\right)}, \quad \forall x,y \in \X^n, \Y^n
\end{equation*}
where $d:\mathcal{X}\times \mathcal{Y} \to \mathbb{R}$ is a single letter decoding
metric. We call such decoders as linear decoders since the decoding metric it computes
is indeed linear in the joint empirical distribution between the codeword and
the received word:
\begin{equation*}
d^{n}\left(x,y\right)=n \sum_{u\in \mathcal{X},v\in\mathcal{Y}}{\hat{P}_{\left(x,y\right)}(u,v)d(u,v)}
\end{equation*}
where $\hat{P}_{\left(x,y\right)}(u,v)$ denotes the joint empirical distribution of $\left(x,y\right)$.
We then say that the linear decoder is induced by the single letter metric $d$.
A significant body of work on the topic of linear decoders exist in the literature. We
refer to \cite{paper:abbe},\cite{paper:Csiszer-Narayan1995}, and the references within,
for a detailed review of such decoders.
Examples of linear decoders are the maximum likelihood (ML) or the maximum a posteriori (MAP) decoders,
whereas the MMI decoder is not linear.

The main advantage of a linear decoder is that, when used with appropriately
structured codes, it allows significant reduction in the decoding complexity.
In this regard, a convenient structure is that of linear encoders, which produce
codewords out of a linear  transformation
$$x = Gu,$$
where $u \in \X^{nR}$ contains the information bits and $G\in \X^{n \times nR} $ is the generating matrix.
With such encoders, linear decoders allow the use of techniques such as the Viterbi algorithm, where significant
complexity reduction is made possible in the optimum decoding (e.g. maximum likelihood sequence decoding (MLSE)) of convolutional codes,
or the message-passing (belief propagation) algorithms adopted for the decoding of several modern coding schemes \cite{book:urbanke}.

The expected reduction in decoding complexity discussed so far is possible only
when the code is appropriately structured. However,
for the proof of existence of linear universal decoders in this paper,
we rely on the random coding argument, where one establish the existence of a deterministic
code which yields good performance, without explicitly constructing the code. For the complexity claims,
one still needs to investigate on whether appropriately structured encoders, can realize
the performance guaranteed by the random coding argument.
However, from an argument of Elias \cite{elias},
we already know that this is possible for binary symmetric channels, where it is sufficient to
consider random binary codes to achieve the performance of arbitrary random codes; this argument
is based on the fact that pairwise independence between the codewords is a
sufficient assumption which has been further generalized in \cite{paper:como:bsc:codes:mismatched}.

A class of decoders slightly more general than linear decoders and
called {\it generalized linear decoders} in \cite{paper:abbe} is defined as follows.
Instead of requiring the condition that the decoding metric is additive, it is only required that the
score function breaks into the {\it maximization} of a finitely many additive
metrics, cf. Definition 2 in \cite{paper:abbe}.
The purpose of studying generalized linear decoders is that all properties
mentioned above for linear decoders still hold for generalized decoders.


\subsection{Linear universal decoders}
In view of constructing universal codes of manageable complexity, a first legitimate question
is to ask wether linear universal decoders can be constructed.
Not surprisingly, some compound channels do not admit a linear universal decoder.
It is known that, when the set $\mathcal{S}$ is convex and compact, the maximum likelihood,
tuned to the channel offering the least mutual information (for the optimal input distribution in (\ref{bbt}))
serves as a compound capacity achieving decoder \cite{paper:Csiszer-Narayan1995}.
In \cite{thesis:abbe,paper:abbe}, it is shown that this result still holds if the set $\Sc$ is non
convex but one-sided, cf. Definition 4 in \cite{paper:abbe}.
Moreover, authors in \cite{paper:abbe} show that if $\Sc$ is a finite union of one-sided sets,
a generalized linear decoder that achieves the compound capacity can be constructed.

In this paper, we construct a generalized linear
decoder which achieve 94\% of the compound capacity on any BMC compound set,
by using the same two metrics, chosen irrespective of the given compound set.

The remainder of this paper is organized as follows. We review known results
for DMC's and then introduce the notations in the next section.
The problem statement is discussed in section \ref{section:problem:statement}.
 We then present the main results for BMC's in section \ref{section:binary:universal}.

\section{Known results for DMC}
We consider discrete memoryless channels with input alphabet $\mathcal{X}$ and output
alphabet $\mathcal{Y}$. 
A DMC is described by a probability transition matrix $W$, each row of which is the
conditional probability distribution of the output $\mathcal{Y}$ given input
$\mathcal{X}$.
We denote by $\Sc$, a compound set of DMC's.
While the set of channels is known to both the transmitter and the receiver,
the exact channel of communication, denoted by $W_0$, is unknown to them.

{
We assume that the transmitter and receiver operate synchronously over blocks of $n$ symbols.
In each block, a message $m\in \left\{1,2,\ldots,2^{nR}\right\}$ is mapped by an encoder
\begin{equation*}
F_n:=\left\{1,2,\ldots,2^{nR}\right\} \to \mathcal{X}^{n}=\{0,1\}^{n}
\end{equation*}
to $F_n(m)=x_m$, referred to as the $m^{th}$ codeword. The receiver
upon observing a word, drawn from the distribution
\begin{equation*}
W^{n}\left(y|x_m\right)=\prod_{i=1}^{n}{W\left(y(i)|x_{m}(i)\right)},
\end{equation*}
applies a decoding map
\begin{equation*}
G_n:\mathcal{Y}^{n} \to \left\{1,2,\ldots,2^{nR}\right\}.
\end{equation*}

The average probability of error, averaged over a given
code $(F_n,G_n)$ 
for a specific
channel $W$, is expressed as
\begin{equation*}
P_{e}\left(F_n,G_n,W\right)=\frac{1}{2^{nR}}\sum_{m=1}^{2^{nR}}{\sum_{y:G_n(y)\ne m}{W^{n}\left(y|x_m\right)}}.
\end{equation*}

We say that a rate $R$ is achievable for a given compound set $\mathcal{S}$, if for any $\epsilon >0$,
there exist a block length $n$ and code 
with rate at least $R$,
such that, for all $W\in \mathcal{S}$, $P_{e}\left(F_n,G_n,W\right)<\epsilon$. The supremum of such
available rates is called the \textit{compound channel capacity}, denoted as $C(\mathcal{S})$. Blackwell et al. \cite{paper:blackwell},
formulated the expression for compound channel capacity as
\begin{equation*}
C(\mathcal{S}) \triangleq \underset{P}{\max}\underset{W\in S}{\inf} I\left(P,W\right).
\end{equation*}

Before proceeding with the problem statement, we will introduce a few notations and some useful definitions.



\begin{definition}(Generalized Linear Decoder)\\
Let $d_1,d_2,\ldots,d_K$ be $K$ single letter metrics, where $K$ is a finite integer. A
generalized linear decoder induced by these metrics is defined by the decoding map:
\begin{eqnarray*}
G_n(y)&=&\arg \underset{m}{\max} \displaystyle \vee_{k=1}^{K} \sum_{i=1}^{n} d_k\left(x_m(i),y(i)\right) \\
&=&\arg \underset{m}{\max} \displaystyle \vee_{k=1}^{K} \mathbb{E}_{\hat{P}_{\left(x_m,y\right)}}[d]
\end{eqnarray*}
where $\vee$ denotes the \textit{maximum} operator and $\hat{P}_{\left(x_m,y\right)}$ denotes the joint empirical distribution of $(x_m,y)$.
\end{definition}
An example of generalized linear decoder is the generalized likelihood ratio test (GLRT)
which, for a given collection of channels $W_1,\ldots, W_K$, is induced by the metrics
\begin{equation*}
\log W_k(v|u),\quad \forall u \in \mathcal{X},v\in \mathcal{Y},  k=1,\ldots,K.
\end{equation*}

We now denote by $P$ an input distribution. 
For a channel denoted by $W_k$, we use $\mu_k$ to denote the joint distribution of an input and output pair through $W_k$, i.e., $\mu_k = P \circ W_k$.
We denote by $\mu_{k}^p$ the product measure of the marginal input distribution $P$ and
the marginal output distribution $(\mu_k)_\Y (y) = \sum_{x \in \X} \mu_k(x,y)$.
Hence,
\begin{eqnarray*}
I\left(P,W_k\right)
&=& D\left(\mu_k \lVert \mu_k^{p}\right).
\end{eqnarray*}

\begin{lemma}\label{monster}
When the true channel is $W_0$ and a generalized linear decoder induced by the
single-letter metrics $\{d_k\}_{k=1}^K$ is used, we can achieve the following rate
\begin{align*}
I_{\text{MIS}}(P_X, W_0, \{d_k\}_{k=1}^K) = \bigvee_{k=1}^K \min_{\mu \in \A_k} D(\mu \| \mu_0^p)  
\end{align*}
where
\begin{eqnarray*}
 \A_k = \{ \mu: \mu^p = \mu_0^p,  \mathbb{E}_\mu[d_k] \geq \vee_{j=1}^K \mathbb{E}_{\mu_0}[d_j] \}, \quad  \forall 1 \leq k \leq K.
\end{eqnarray*}
\end{lemma}
In particular, if $k=1$, this is the mismatched result in \cite{paper:Csiszer-Narayan1995} and \cite{paper:Merhav1994},
and if $k \geq 2$, it is a consequence of \cite{paper:Csiszer-Narayan1995},\cite{paper:Merhav1994}, as discussed in \cite{paper:abbe}.
Extensive coverage of the \textit{mismatched problem} \cite{paper:Lapidoth-Narayan1998} appears
in the literature \cite{paper:mismatch:balakirsky},\cite{paper:mismatch:balakirsky2},\cite{paper:Csiszer-Narayan1995},\cite{paper:mismatch:Hui},\cite{paper:mismatch:Lapidoth2},\cite{paper:Merhav1994} as well as \cite{paper:abbe}.

\begin{definition}[one-sided sets]
A set $\Sc$ of DMC's is one-sided with respect to an input distribution $P$,
if $W_{\Sc} =\arg \underset{W\in cl(\Sc)}{\min} I \left(P,W\right)$ is unique and if
$$D\left(\mu \lVert \mu_{S}^{p}\right) \ge D\left(\mu \lVert \mu_{\Sc}\right)+ D\left(\mu_{\Sc} \lVert \mu_{\Sc}^{p}\right)$$
for any $\mu= P \circ W$, where $W \in \Sc$ and $\mu_\Sc = P \circ W_\Sc$.
A set $\Sc$ of DMCs is a union of one-sided sets if $\Sc= \cup_{k=1}^K \Sc_k$ for some $K <\infty$ and if the $\Sc_k$'s are one-sided with respect to $P_* = \arg \underset{P}{\max} \underset{W \in \Sc}{\inf} I(P, W)$.
\end{definition}

\begin{theorem}\label{ludthm}
For a compound set $\Sc=\displaystyle \bigcup_{k=1}^{K} \Sc_{k}$ which is a union of one-sided sets,
the generalized linear decoder induced by the metrics
\begin{equation*}
d_k=\log \frac{W_{\Sc_k}}{\left(\mu_{\Sc_k}\right)_{Y}}, \quad 1\le k\le K
\end{equation*}
where $W_{\Sc_k} = \arg \underset{W\in cl(\Sc_k)}{\min} I \left(P_*,W\right)$,
is compound capacity achieving.
Moreover, if the true channel $W_0$ is known to belong to $\Sc_k$, this decoder allows to achieve
the rate,
\begin{align}
I_{\text{MIS}}(P_*, W_0, \log W_{\Sc_k}).
\end{align}
\end{theorem}

Notice that, this decoder requires the full knowledge of the compound set.
In other words, knowing the compound capacity as well as the compound capacity achieving input
distribution (which for instance, suffices for MMI) will not suffice to construct such a decoder.
One can interpret this decoder as follows: the MMI decoder allows to achieve compound capacity
on any compound set, by taking a ``degenerate generalized linear'' decoder with infinitely (and uncountably) many metrics: the a posteriori metrics of all
possible DMCs with the given alphabets.
Hence, there is no linear property (and consequent benefit) for such a decoder.
However, what Theorem \ref{ludthm} says, is that, since we have the knowledge of the compound set,
we can use it to tune a generalized linear decoder which will still achieve
compound capacity by picking only the important channels and corresponding a posteriori metrics.
Our goal in this paper is to investigate whether further simplification in the above generalized linear
 decoder can be made when restricted to BMC's. More specifically, we address the
 possibility of building a universal decoder tuned to metrics, chosen
independently to the given compound BMC set.

Using the symmetry properties occurring in the BMC's, we have the following result (cf. \cite{thesis:abbe}).

\begin{lemma}\label{hint}
Let $P_1,P_2$ be two stochastic matrices of size $2 \times 2$, such that $\det (P_1P_2) >0$.
Let $C$ a set of binary vectors of length $n$ with fixed composition. For any $x_1,x_2 \in C$ and $y \in \{0,1\}^n$, we respectively have
$$P_1(y|x_1)\, >(=) \,P_1(y|x_2) \, \Rightarrow \, P_2(y|x_1)\, > (=)\, P_2(y|x_2).$$
\end{lemma}

\noindent
Note that the condition $\det (P_1P_2) >0$ simply means that $P_1$ and $P_2$ have their maximal value within a column at the same place.


Hence, we would like to investigate whereas for BMC, the generalized linear decoder of Theorem \ref{ludthm} can be simplified, so as to require only few metrics and independently of the compound set.

\section{Problem statement}\label{section:problem:statement}

\subsection{The $\alpha$ and $\beta$ game}
From now on, we only consider BMC's.

Let the parameter $\alpha$ be defined as
\begin{align}
\alpha= \max_{P}\inf_{W \in BMC}{\frac{I\left(P,W\right)}{C(W)}}. \label{alpha}
\end{align}
and the distribution $P$ which achieve $\alpha$ is denoted by $P_{opt}$, i.e.,
\begin{equation}\label{eq:Popt}
P_{\text{opt}} := \arg \max_{P}\inf_{W \in BMC}{\frac{I\left(P,W\right)}{C(W)}}.
\end{equation}

The term $\alpha$ is the ratio of the maximum achievable rate to the channel capacity, for the worst
possible channel in the compound set, when a single input distribution is chosen.

Let $K \in \mZ_+$ and let $I_{\text{mis}}\left(P_{opt},W_{0},\left\{ d_{k} \right\}_{1}^{K} \right)$
be the achievable mismatched rate on a channel $W_0$, using a generalized linear decoder
induced by the $K$ metrics $\left\{ d_{k} \right\}_{1}^{K}$. The expression
of $I_{\text{mis}}\left(P_{opt},W_{0},\left\{ d_{k} \right\}_{1}^{K} \right)$ is given
in the next section, and is proved to be an achievable rate
in \cite{paper:Csiszer-Narayan1995},\cite{paper:Merhav1994} as well as discussed in \cite{paper:abbe}.
Indeed, since we are working with binary input (and output) channels, this
mismatched achievable rate is equal to the mismatched capacity.
We then define another parameter $\beta_K$ by
\begin{align}
\beta_K := \inf_{W_0 \in BMC}{\frac{I_{\text{MIS}}\left(P_{opt},W_{0},\left\{ d_{k} \right\}_{1}^{K} \right)}{I\left(P_{opt},W_0\right)}}. \label{beta}
\end{align}

Clearly $0\le \alpha,\beta \le 1$.
The problem of finding $\alpha$ has already been solved in \cite{paper:shulman}, and the
answer is $\alpha \approx 0.942$, as reviewed below. From Theorem \ref{ludthm} (second part), one
can show that by taking a large enough $K$, we can make $\beta_K$ arbitrarily close to $1$.
Indeed, this relates to the fact that we can approximate the set of DMC's by a covering of
one-sided components (for the uniform input distribution). Hence, one can study the speed
convergence of $\beta_K$, in $K$, to deduce how many metrics in magnitude need to be used
to achieve a given performance. We believe this is an interesting problem, as it captures
the cost (in the number of additive metrics) needed to ``replace'' the MMI decoder with a
generalized linear decoder; and this problem can be addressed for any alphabet dimensions.
However, as motivated in previous section with Proposition \ref{hint}, we hope to get
exactly $\beta_2=1$ for the binary alphabets setting, in which case we do not need to
investigate the speed convergence problem (this will indeed be the case).


\section{Results}\label{section:binary:universal}

\subsection{Optimal input distribution}

The optimization problem for $\alpha$ has been investigated in \cite{paper:shulman} with the following answer.
\begin{theorem}[Shulman and Feder]
$\alpha \approx 0.942$ and $P_{\text{opt}}$ is the uniform distribution.
\end{theorem}

The authors also identified the worst channel to be a Z-channel.
This result is also a ramification of the fact that with uniform source distribution,
 the maximum loss for any channel is less than $5.8\%$
of capacity, as originally reported by Majani and Rumsey \cite{paper:majani}.

\subsection{Optimal Generalized Linear Decoder}



We represent a BMC by a point in $(a,b) \in [0,1]^2$, with the following mapping to specify the BMC
\begin{equation*}
\begin{pmatrix} a & 1-a \\ 1-b & b\end{pmatrix},\quad 0\le a,b \le 1.
\end{equation*}
\begin{definition}
Let $B^- =\{(a,b)  \in [0,1]^2 | a+b < 1\}$ and $B^+ =\{(a,b)  \in [0,1]^2 | a+b \geq 1\}$,
and let $U$ denote the uniform binary distribution.
\end{definition}
Note that $B^-$ parameterizes the set of BMC's which are {\it flipping-like},
in the sense that, assuming the input and output alphabets to be $\{0,1\}$, for any
given output $y$ of a BMC in $B^-$, it is more likely that the sent input is $1+y$ (mod 2).
Similarly, $B^+$ parameterizes the set of BMC's which are {\it non-flipping-like},
containing in particular the set of channels for which $a+b=1$, which are all the
pure noise channels (zero mutual information).

\begin{prop}\label{l1}
For any $i \in \{-,+\}$ and any $W_0 \in B^i$, we have
\begin{align*}
I_{\text{MIS}} \left(U, W_0,\log W_1\right) =
     \begin{cases}
     I\left(U,W_0\right)    & \text{if } W_1 \in B^i \\
        0  & \text{otherwise.}
     \end{cases}
\end{align*}
\end{prop}
This proposition tells us that, as long as the channel used for the decoding ($W_1$)
is in the same class ($B^-$ or $B^+$) as the true channel ($W_0$), the mismatched
mutual information is equal to the mutual information itself; i.e., equal to all
of the mutual information being evaluated with the uniform input distribution.
If instead the channel and the metrics each other hail from different class,
then the mismatched mutual information is zero.
\begin{IEEEproof}
As defined in Proposition \eqref{monster}, we have
\begin{align}
I_{\text{MIS}} \left(U, W_0, \log W_1\right) = \inf_{\mu \in \A} D\left(\mu \| \mu_0^p\right) \label{double}
\end{align}
where $\A= \{\mu: \mu^p = \mu_0^p,  \mathbb{E}_\mu \log W_1 \geq \mathbb{E}_{\mu_0} \log W_1 \}$.
Note that the channels $W$ which induce a $\mu$ such that
$\mu^p = \mu_0^p$ is parameterized in $[0,1]^2$
by the line passing through $\mu_0$ with a slope of 1. Hence,
since $\mu_0 \in \partial \A$ (the boundary of $\A$), it is easy to verify that
the region $\A$ is the segment starting at $\mu_0$ and going either up or
down (with slope $1$). This leads to two possibilities, either $\mu_0^p \in \A$
and \eqref{double} is 0, or $\mu_0^p \notin \A$ and the minimizer
of \eqref{double} is $\mu_0$, implying the claims of Proposition \ref{l1}.
\end{IEEEproof}

\begin{prop}
For any binary input/output channel $W_0$ and for any binary symmetric
channel $W_1$, i.e., $W_1(0|0)=W_1(1|1)$, we have
\begin{align*}
I_{\text{MIS}} \left(U, W_0,\{\log W_1, \log \ww_1 \}\right) = I\left(U,W_0\right),
\end{align*}
where $\ww_1$ is the BSC defined by $\ww_1(0|0)=1-W_1(0|0)$.
\end{prop}

\begin{IEEEproof}
As defined in Lemma \eqref{monster}, we have
\begin{align}
I_{\text{MIS}} \left(U, W_0,\{ \log W_1, \log \ww_1\}\right) = \bigvee_{k=1}^2 \inf_{\mu \in \A_k} D\left(\mu \| \mu_0^p\right) , \label{targ}
\end{align}
where $\A_k = \{  \mu: \mu^p = \mu_0^p,  \mathbb{E}_\mu \log W_k \geq \vee_{j=1}^2 \mathbb{E}_{\mu_0} \log W_j \}$, $k=1,2$, where $W_2=\ww_1$.
Note that, although in general taking the likelihood metrics as opposed to the a posteriori metrics makes an important difference when defining a generalized linear decoder (cf. \cite{paper:abbe}), here it does not, since we are working with BSC channels for the metrics.
Assume w.l.o.g. that $W_0, W_1 \in B^-$. Then, a straightforward computation shows that
$$\bigvee_{j=1}^2 \mathbb{E}_{\mu_0} \log W_j = \mathbb{E}_{\mu_0} \log W_1.$$
Hence
$$ \inf_{\mu \in \A_1} D\left(\mu \| \mu_0^p\right) = I_{\text{MIS}} \left(U, W_0, \log W_1\right),$$
and from Proposition \ref{l1}
$$ I_{\text{MIS}} \left(U, W_0, \log W_1\right) = I\left(U, W_0\right).$$
Moreover, note that for any channel $W_0$, if we define $\ww_0$ to be the reverse BSC (cf. Figure \ref{fig:dmc_proof1}), and $\mu_0$, $\widetilde{\mu}_0$ to be the corresponding measures, we have $\mu_0^p = \widetilde{\mu}_0^p$, $$\mathbb{E}_{\mu_0} \log W_1=\mathbb{E}_{\widetilde{\mu}_0} \log \ww_1$$ and
\begin{align*}
\A_2&=\{ \mu: \mu^p = \mu_0^p, \mathbb{E}_{\mu} \log W_2 = \mathbb{E}_{\mu_0} \log W_1 \} \\
&=\{ \mu: \mu^p = \widetilde{\mu}_0^p, \mathbb{E}_{\mu} \log \ww_1 = \mathbb{E}_{\widetilde{\mu}_0} \log \ww_1 \}.
\end{align*}
Therefore,
\begin{align*}
\inf_{\mu \in \A_2} D(\mu \| \mu_0^p) = I_{\text{MIS}} \left(U,\ww_0, \ww_1\right)=I_{\text{MIS}} \left(U, W_0, W_1\right),
\end{align*}
and both terms in the RHS of \eqref{targ} are equal to $I\left(U, W_0\right)$.
\begin{figure}[h]
	\begin{center}
	\includegraphics[scale=1.2]{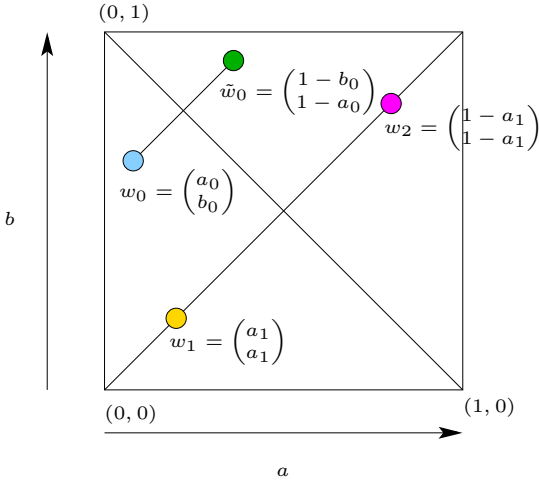}
	\caption{Reverse channels in the BMC setting}
	\label{fig:dmc_proof1}
	\end{center}
	\end{figure}
\end{IEEEproof}

An extended discussion and alternate proofs of previous results can be found in \cite{thesis:ratna}.
\begin{corol}
We have $\beta_2=1$, which is achieved by picking any two metrics of the
form $d_1 = \log W_1$, $d_2 = \log \ww_1$, as long as $W_1$ is a BSC (and $\ww_1$ its reverse BSC).
\end{corol}

\begin{corol}
For any compound sets $\Sc$, 94.2\% of the compound capacity can be achieved by using a
generalized linear decoder induced by two metrics. Moreover, if the optimal input
distribution (achieving compound capacity) is uniform, we can achieve 100\% of the
compound capacity with two metrics.
\end{corol}

Note: if the optimal input distribution is non-uniform, it may still be
possible to achieve 100\% of the compound capacity with two metrics, but the above
results will have to be adapted to the non-uniform input distribution case.

\section{Discussion}\label{section:conclusion}
In this paper, we have shown that, for binary input binary output memoryless
channels, compound capacity achieving decoders can have a much
simpler structure than the Maximum Mutual Information (MMI) decoder.
These decoders, namely the generalized linear decoders, preserve
many features of the MMI decoder.
When the input distribution is chosen to be uniform, a generalized
linear decoder, using channel independent metrics (i.e., the metrics are selected without being aware
of the channel rule) is shown to
achieve the same rate as that of an optimum decoder (based on the exact channel rule).
On the other hand, for any arbitrary compound BMC, at least 94.2\% of the compound capacity can be realized
by such a decoder.
%
%
Finally, a natural extension of this work would be to investigate the case of non-binary
alphabets. Even for binary input and ternary outputs, it does not seem straightforward to
establish whether $\beta_K$ can be made exactly $1$, for $K$ large enough (although one can show
that it must tend to $1$ using results from \cite{paper:abbe}).








\begin{thebibliography}{}

\end{thebibliography}


\begin{thebibliography}{77}

\bibitem{thesis:abbe}
E.~Abbe, ``Local to geometric Methods in Information Theory, PhD thesis," Massachusetts Institute of Technology, 2008.
\bibitem{paper:abbe}
E.~Abbe and L.~Zheng, ``Linear Universal Decoding for Compound Channels: a Local to Global Geometric Approach," {\tt arXiv:0809.5217v1  [cs.IT]}, Sep. 2008.

\bibitem{paper:mismatch:balakirsky}
V. B.~Balakirsky, ``Coding theorem for discrete memoryless channels with given decision rules," in Proc. 1st French–Soviet Workshop on
Algebraic Coding (Lecture Notes in Computer Science 573), G.~Cohen, S.~Litsyn, A.~Lobstein, and G.~Z´emor, Eds. Berlin, Germany: Springer-
Verlag, pp. 142-150, Jul. 1991.
\bibitem{paper:mismatch:balakirsky2}
V. B.~Balakirsky, ``A converse coding theorem for mismatched decoding at the output of binary-input memoryless channels," IEEE Trans. Inform.
Theory, vol. 41, pp. 1889–1902, Nov. 1995.

\bibitem{paper:blackwell}
D.~Blackwell, L.~Breiman and A.J.~Thomasian, ``The Capacity of a class of channels," \textit{Ann. Mathe.Stat}., vol. 30, pp. 1229-1241, Dec. 1959.

\bibitem{paper:como:bsc:codes:mismatched}
G.~Como and F.~Fagnani, ``The capacity of abelian group codes over symmetric channels," IEEE Trans. Inform. Theory, submitted

\bibitem{paper:cover:broadcast}
T. M.~Cover, ``Broadcast channels," IEEE Trans. Inform. Theory, vol. IT-18, pp. 2–14, Jan. 1972.
\bibitem{book:Csiszar}
I.~Csiszar and J.~Korner, \textit{Information Theory:Coding Theorems for Discrete Memoryless Systems}. New York: Academic, 1981.
\bibitem{paper:Csiszer-Narayan1995}
I.~Csiszer, P.~Narayan, ``Channel Capacity for a Given Decoding Metric", IEEE Trans. Inform. Theory, vol. 41, no.1, Jan. 1995.

\bibitem{paper:compound:Dobrushin}
R. L.~Dobrushin, ``Optimum information transmission through a channel with unknown parameters," Radio Eng. Electron., vol. 4, no. 12, pp.
1–8, 1959.

\bibitem{elias}
P.~Elias, ``Coding for two noisy channels", in Proc. 3rd London Symp. Information Theory, London, U.K., pp. 61-76, 1955.

\bibitem{paper:feder-lapodoth}
M.~Feder and A.~Lapidoth, ``Universal decoding for channels with memory," IEEE Trans. Inform. Theory, vol. 44: pp. 1726-1745, Sep. 1998.

\bibitem{paper:Goppa}
V.~Goppa, ``Nonprobabilistic mutual information without memory," Problems of Control and Information Theory, vol. 4, pp. 97-102, 1975.


\bibitem{paper:mismatch:Hui}
J. Y. N.~Hui, ``Fundamental issues of multiple accessing," Ph.D. Thesis, Mass. Inst. Technol., Cambridge, MA, 1983.


\bibitem{paper:mismatch:Lapidoth2}
A.~Lapidoth, ``Mismatched decoding and the multiple-access channel," IEEE Trans. Inform. Theory, vol. 42, pp. 1439–1452, Sep. 1996.


\bibitem{paper:Lapidoth-Narayan1998}
A.~Lapidoth and P.~Narayan, ``Reliable Communication under channel uncertainty," IEEE Trans. Inform. Theory, vol.40, No.10, pp:2148-2177, Oct. 1998.
\bibitem{paper:lapidoth-telatar1998}
A.~Lapidoth and I.E.~Telatar, ``The compound channel capacity of a class of finite state channels," IEEE Trans. Inform. Theory, Vol.44,No.3,May 1998.
\bibitem{paper:majani}
E.E.~Majani and H.~Rumsey, ``Two results on binary-input discrete memoryless channels," In Proc. IEEE Int. Symp. Information Theory,
Budapest, Hungary, pp. 104, 1991.
\bibitem{paper:Merhav1994}
N.~Merhav, G.~Kaplan, A.~Lapidoth and S.~Shamai, ``On information rates for mismatched decoders," IEEE Trans. Inform. Theory, vol 40. No.6 pp. 953-1967, Nov. 1994.
\bibitem{thesis:ratna}
R.~Pulikkoonattu, ``Linear Universal Decoder for Compound Discrete Memoryless Channels," MS Thesis, EPFL, Jul. 2009.
\bibitem{book:urbanke}
T.~Richardson and R.~Urbanke, \textit{Modern Coding theory}, Cambridge University Press, 2007.
\bibitem{paper:shulman}
N.~Shulman and M.~Feder, ``The uniform Distribution as a Universal Prior," IEEE Trans. Inform. Theory, vol.50.No.6, Jun. 2004.
\bibitem{paper:ziv}
J.~Ziv, ``Universal decoding for finite state channels," IEEE Trans. Inform. Theory, vol. 31, pp. 453-460, Jul. 1985.











\end{thebibliography}
\bibliographystyle{plain}

\end{document}